\documentclass[twocolumn,secnumarabic,amssymb, nobibnotes, aps, superscriptaddress]{revtex4-1}

\usepackage{graphicx} 
\usepackage{amsmath}
\usepackage{float}
\usepackage{color}
\usepackage[english]{babel}
\usepackage{array}
\usepackage[T1]{fontenc}
\usepackage[usenames,dvipsnames]{xcolor}
\usepackage{setspace}
\usepackage{hyperref}
\usepackage{bm}
\begin{document}
\title{Symmetrization of Thin free-standing Liquid Films via Capillary-Driven Flow}
\author{Vincent Bertin}
\thanks{These two authors contributed equally}
\affiliation{Univ. Bordeaux, CNRS, LOMA, UMR 5798, 33405 Talence, France.}
\affiliation{UMR CNRS Gulliver 7083, ESPCI Paris, PSL Research University, 75005 Paris, France.}
\author{John Niven}
\thanks{These two authors contributed equally}
\affiliation{Department of Physics \& Astronomy, McMaster University, Hamilton, Ontario L8S 4M1, Canada.}
\author{Howard A. Stone}
\affiliation{Department of Mechanical and Aerospace Engineering,
Princeton University, Princeton, New Jersey 08544, USA}
\author{Thomas Salez}
\affiliation{Univ. Bordeaux, CNRS, LOMA, UMR 5798, 33405 Talence, France.}
\affiliation{Global Station for Soft Matter, Global Institution for Collaborative Research and Education, Hokkaido University, Sapporo, Hokkaido 060-0808, Japan.}
\author{Elie Rapha\"{e}l}
\affiliation{UMR CNRS Gulliver 7083, ESPCI Paris, PSL Research University, 75005 Paris, France.}
\author{Kari Dalnoki-Veress}
\email{dalnoki@mcmaster.ca}
\affiliation{UMR CNRS Gulliver 7083, ESPCI Paris, PSL Research University, 75005 Paris, France.}
\affiliation{Department of Physics \& Astronomy, McMaster University, Hamilton, Ontario L8S 4M1, Canada.}
%\date{\today}
\begin{abstract}
We present experiments to study the relaxation of a nano-scale cylindrical perturbation at one of the two interfaces of a thin viscous free-standing polymeric film. Driven by capillarity, the film flows and evolves towards equilibrium by first symmetrizing the perturbation between the two interfaces, and eventually broadening the perturbation. A full-Stokes hydrodynamic model is presented which accounts for both the vertical and lateral flows, and which highlights the symmetry in the system. The symmetrization time is found to depend on the membrane thickness, surface tension, and viscosity.
\end{abstract}
\maketitle

Surface tension will smooth out small interfacial perturbations on a thin liquid film, as the curvature of the perturbation profile induces a Laplace pressure that drives a viscous flow. This capillary-driven levelling causes the brush strokes on paint to flatten, or the spray of small droplets to result in a uniform film. Such flows have been studied in great detail and much of the theoretical framework is provided by the lubrication approximation, whereby one can assume that the flow in the plane of the film dominates, and that the velocity vanishes at the solid-liquid interface~\cite{oron1997long,craster2009dynamics}. 
In contrast, for a free-standing liquid film there is no shear-stress at both liquid-air interfaces which modifies the boundary conditions and results in a different phenomenology~\cite{oron1997long}. These boundary conditions can arise in a variety of situations such as biological membranes \cite{tanaka2005polymer}, soap films \cite{couder1989hydrodynamics, rutgers1996,  aradian2001, georgiev2002, huang2004, seiwert2013}, liquid-crystal films~\cite{stannarius2006, harth2011, eremin2011}, fragmentation processes~\cite{villermaux2007fragmentation}, or energy-harvesting technologies~\cite{nie2019power}.
\begin{figure*}
\centering
\includegraphics[width=0.85\textwidth]{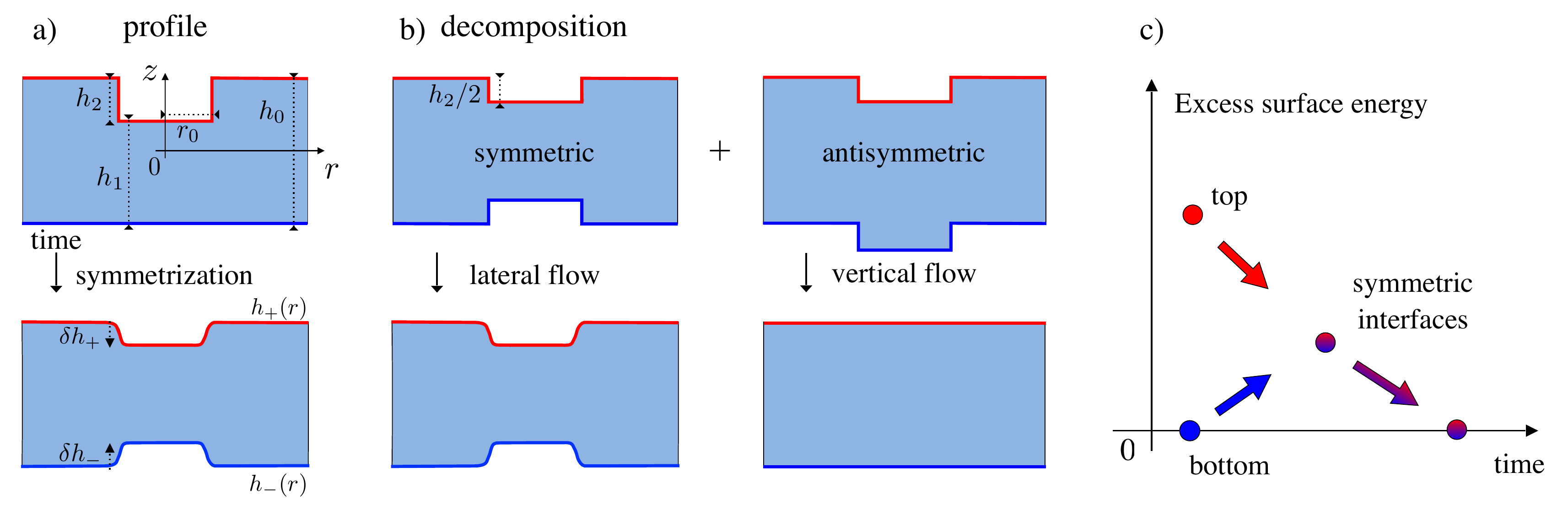}
\caption{(a) Schematic of an initial cylindrical hole of depth $h_2$ and radius $r_0$, on one side of a polystyrene free-standing film, which evolves towards symmetric. (b) Symmetric-asymmetric decomposition of the interfacial profiles. A symmetric profile leads to lateral flow, while an antisymmetric one leads to vertical flow. (c) Schematic of the evolution of the excess surface energy. The top and bottom surface energies equalize rapidly before vanishing in tandem on larger time scales. }
\label{fig:schem}
\end{figure*} 

The dynamics of liquid sheets has been studied in great detail in the past decades~\cite{howell1996models,ribe2002general}, and shows similarities with the mechanics of elastic plates. The evolution can be described with two dominant modes, which are the stretching and bending modes associated with momentum and torque balances. At macroscopic scales, a viscous sheet experiences bending instabilities such as wrinkling~ \cite{taylor1969instability, debregeas1998life, da2000rippling,boudaoud2001singular}, and folding~ \cite{ribe2003periodic} when submitted to compressive forces. Such viscous buckling phenomena occur in various contexts, like tectonic-plate dynamics~\cite{biot1961theory,guillou1995laboratory} and industrial float-glass processes~\cite{pearson1985mechanics, pfingstag2011linear, srinivasan2017wrinkling}.  

In thin free-standing films, surface tension is dominant and stabilizes the interfaces against buckling~\cite{howell1996models}. Most theoretical models in this context assume that the interfaces are mirror-symmetric, and thus focus on the stretching mode, also called the symmetric mode. This approach is employed to study the rupture dynamics of films in the presence of disjoining forces that destabilize long waves in thin film~\cite{ruckenstein1974spontaneous, prevost1986nonlinear, erneux1993nonlinear, vaynblat2001rupture, matar2002nonlinear, thete2016self, ilton2016, choudhury2019role}. Recently, using nanometric free-standing polystyrene (PS) films, Ilton \emph{et al.} observed that a film with initially asymmetric interfaces symmetrized over short time scales \citep{MI2016}. This symmetrization was attributed to flow perpendicular to the film, but the dynamics was not accessible experimentally. 

In this Letter we study the viscocapillary relaxation dynamics of a nanoscale cylindrical perturbation initially present on one of the two interfaces of a thin free-standing PS film. Both the symmetric (viscous stretching) and antisymmetric (viscous bending) modes are probed with experiments (see Fig.~\ref{fig:schem}a-b). Atomic force microscopy (AFM) is used to obtain the profiles of the top and bottom interfaces. A full-Stokes flow linear hydrodynamic model is developed to characterize the relaxation dynamics of the two modes. To provide an intuitive understanding of the energy dissipation as the film relaxes, we turn to the schematic plot of the excess surface energy as a function of time, shown in Fig.~\ref{fig:schem}(c). Initially, the top interfacial profile, denoted $h_+$, has a high excess energy due to the additional interface that forms the hole, while the bottom interfacial profile $h_-$ is flat and hence has no excess surface energy. The excess free energy resulting from the perturbation drives a flow that is mediated by viscosity, $\eta$. As the film evolves, the total energy dissipates as the excess interface decreases. Apart from that global energy dissipation, the symmetrization process requires some energy transfer from the top interface to the bottom one -- a coupling that is dominated by vertical flow. Once both interfaces are mirror-symmetric, they relax in tandem dominated by lateral flow. Remarkably, the temporal evolution of the interfacial profiles, when appropriately decomposed into their symmetric and anti-symmetric components is found to obey power laws.

Thin films of PS are prepared using a method similar to that previously described \cite{MB2014, MI2016}. PS with molecular weight $M_{\text{w}}$ = 183 kg/mol (Polymer Source Inc., polydispersity index = 1.06) is dissolved in toluene (Fisher Scientific, Optima grade) with concentrations of 2 \% and 7.5 \% by weight. Thin films are prepared by spin coating from solution onto freshly cleaved mica sheets (Ted Pella), and annealed at 130 $^{\circ}$C in vacuum (1$\times10^{-5}$ mbar) for 24 hours. The films have thicknesses $h_1=530$~nm and $h_2=80$~nm, as measured using ellipsometry (Accurion, EP3). The free-standing films are then prepared in a two-step process inspired by the work of Backholm \emph{et al.} \cite{MB2014}. Films are floated from the mica substrates, onto the surface of ultrapure water (18.2 M$\Omega\cdot$cm) and picked up on a thin circular stainless steel washer (thickness = 0.1 mm, AccuGroup, UK), creating a free-standing thin film supported only at the edges of the washer. The thicker film, with $h_1$ = 530 nm, is picked up on a washer with an internal diameter of 3 mm, and heated above the glass-transition temperature of PS, $T_{\text{g}} \approx 100~^\circ$C, on a hot stage (Linkam, UK). Similarly, the thinner film with $h_2$ = 80 nm, is transferred from the water to a washer with an internal diameter of 5 mm. This film is rapidly heated (100 $^{\circ}$C/min) on a hot stage to 125 $^{\circ}$C for several seconds under the view of an optical microscope. During the heating, cylindrical holes are nucleated in the film, and their radii grow exponentially with time \cite{ilton2016,GD1995, KDV1999, CR2006}. When the holes become visible, the film is quenched to room temperature, resulting in a free-standing film with holes of diameter 1 -- 10 $\mu$m randomly distributed throughout. The two films are then placed in direct contact and adhere through van der Waals forces, and the larger diameter washer can be removed. This process results in a free-standing film of thickness $h_0 = h_1 + h_2$, with cylindrical holes of depth $h_2$ [see Fig.~\ref{fig:schem}(a, top)]. 

\begin{figure}
    \includegraphics[width=1\columnwidth]{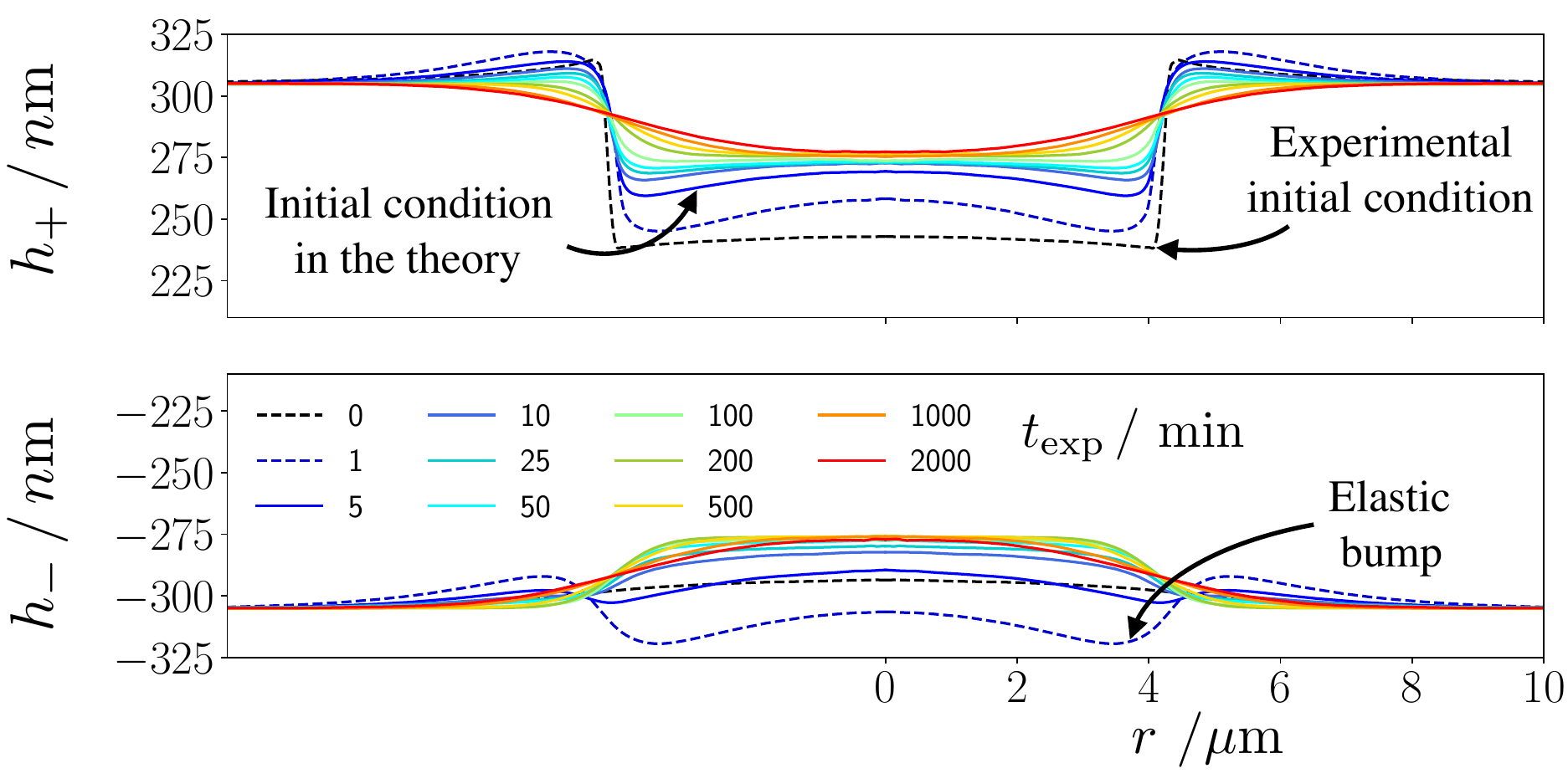}
      \caption{AFM profiles of the top and bottom interfaces of a free-standing hole with $h_2$ = 80~nm, $h_0$ = 610~nm, and $r_0$ = 4.2~$\mu$m (see Fig.~\ref{fig:schem}), at various annealing times $t_{\textrm{exp}}$ as indicated.  An ``elastic bump'' is seen at $t_\textrm{exp}$ = 1 min due to the residual stresses in the film from the sample preparation. The viscous model takes the profiles at 5 min as initial profiles, in order to ignore any prior elastic effect.}
        \label{fig:profs_topbot}
  \end{figure}

The free-standing films are annealed on a hot stage at $T$ = 130 $^{\circ}$C and covered with a glass coverslip to ensure a uniform temperature. After a given amount of annealing time the film is temporarily quenched to room temperature, thus returning to the glassy state where flow becomes arrested. The surface profiles of three holes in the same film are then measured after each annealing step using AFM (Bruker Multimode). Since the film is free-standing and has two polymer-air interfaces, both the top and bottom profiles can be measured. 
%Imaging the bottom side can be accomplished by simply flipping the sample, since the thickness of the washer is small enough not to interfere with the AFM cantilever. 
The angular-averaged profiles of the top and bottom interfaces are extracted at each time step to provide a cross-section of the film as it evolves, as shown in Fig.~\ref{fig:profs_topbot}.
  
Initially, in the region of the hole, the film has significantly different curvature gradients at the top and bottom interfaces, resulting in pressure gradients in both the vertical and lateral directions throughout the film. The initial response of the film in the region of the hole is for the bottom interface to buckle downward, forming a small ($\sim 10$~nm) ``elastic bump''. This feature is not a result of the simple polymeric viscoelastic response to interfacial forces~\cite{Benzaquen2014}, as this response would rather generate an opposite inward motion. Instead, while still being related to a viscoelastic process, it is likely a short-term experimental artifact due to the residual stresses associated with sample preparation~\cite{KDV1999, CR2006}. As the film is annealed further, the elastic bump relaxes on a time scale $\sim 5$ min, which is on the order of the macromolecular relaxation time scale for PS (the reptation time for the PS at the given temperature is $\sim$ 13 min~\cite{AB2003}). 

After relaxation of the elastic bump, the flow results from capillarity and viscosity only. First, there is vertical flow to equilibrate the Laplace pressures of the two interfaces, which results in the symmetrization process. Indeed, two symmetric interfacial profiles at the top and bottom of the film are observed at times larger than $\sim 200$ min. Subsequently, the symmetrized interfacial profiles evolve jointly through lateral uniform flow in order to dissipate the excess surface energy~\cite{erneux1993nonlinear}. The film is annealed for $\sim$ 2000 min before rupturing.

We now turn to a theoretical description. The polymer is assumed to be a Newtonian fluid with viscosity $\eta$. Given the axial symmetry of the problem, we introduce cylindrical coordinates $(r,z)$, as well as the Hankel transforms~\cite{gaskill1978linear} of the velocity field $\vec{u}(r,z,t) = (u_r, u_z)$, and of the interfacial profiles $h_\pm(r,t)$: $\tilde{u}_r(k,z,t) = \int_0^\infty \textrm{d}r \, r \, u_r(r,z,t) \, J_1(kr)$ ,
$\tilde{u}_z(k,z,t) = \int_0^\infty \textrm{d}r \, r \, u_z(r,z,t) \, J_0(kr)$, and
$\tilde{h}_\pm(k,t) = \int_0^\infty \textrm{d}r \, r \, h_\pm(r,t) \, J_0(kr)$,
%\begin{subequations}
%\begin{equation}
%\tilde{u}_r(k,z,t) = \int_0^\infty \textrm{d}r \, r \, u_r(r,z,t) \, J_1(kr)\ ,
%\end{equation}
%\begin{equation}
%\tilde{u}_z(k,z,t) = \int_0^\infty \textrm{d}r \, r \, u_z(r,z,t) \, J_0(kr)\ ,
%\end{equation}
%\begin{equation}
%\tilde{h}_\pm(k,t) = \int_0^\infty \textrm{d}r \, r \, h_\pm(r,t) \, J_0(kr)\ ,
%\end{equation}
%\end{subequations}
where $t$ is time, and the $J_i$ are the Bessel functions of the first kind with indices $i=0,1$. Injecting these forms into the steady Stokes equations, we find: $\partial^3_z \tilde{u}_r + k \partial^2_z \tilde{u}_z - k^2\partial_z \tilde{u}_r - k^3 \tilde{u}_z = 0$ and $\partial_z \tilde{u}_z + k  \tilde{u}_r = 0$,
%\begin{subequations}
%\begin{equation}
%\partial^3_z \tilde{u}_r + k \partial^2_z \tilde{u}_z - k^2\partial_z \tilde{u}_r - k^3 \tilde{u}_z = 0\ ,
%\end{equation}
%and:
%\begin{equation}
%\partial_z \tilde{u}_z + k  \tilde{u}_r = 0\ ,
%\end{equation}
%\end{subequations}
which result in the general solution: 
\begin{subequations} 
\begin{equation}
\begin{split}
\tilde{u}_r =& -\frac{1}{k}\bigg( k A + k z C + D\bigg)  \sinh(kz)\  \\ 
							& - \frac{1}{k}\bigg(k B + k z D + C\bigg) \cosh(kz)\ , 
\end{split}
\end{equation}
\begin{equation}
\tilde{u}_z =\bigg( A+ z C \bigg)  \cosh(kz) + \bigg(B +  z D\bigg)  \sinh(kz)\ ,
\end{equation}
 \end{subequations}
where $A(t)$, $B(t)$, $C(t)$ and $D(t)$ are integration constants. The depth of the hole is assumed to be small in comparison with the total thickness of the film, which is valid for the experiments, so that we can linearize the problem by writing the profiles as $h_\pm = \pm h_0/2 + \delta h_\pm$, where the perturbations $\delta h_\pm$ are small compared to the film thickness $h_0$ at rest. We assume no-shear-stress boundary conditions at both fluid-air interfaces, and neglect the nonlinearities from the scalar projections of the normal and tangential vectors to the interface, which gives:  
\begin{subequations}
\begin{equation}
\label{eq:normal_stress}
\begin{split}
(\pm kA + C \frac{kh_0}{2})\sinh(\frac{kh_0}{2}) \\ + (kB \pm D \frac{kh_0}{2})\cosh(\frac{kh_0}{2}) = \pm \frac{\gamma k^2 }{2 \eta}  \tilde{\delta h}_\pm,
\end{split}
\end{equation}
\begin{equation}
\begin{split}
\left(kA \pm C \frac{kh_0}{2} + D\right)\cosh\left(\frac{kh_0}{2}\right) \\ + \left(\pm kB + D \frac{kh_0}{2} \pm C\right)\sinh\left(\frac{kh_0}{2}\right)  = 0\ ,
\end{split}
\end{equation}
\end{subequations}
where $\gamma$ is the fluid-air interfacial tension. Finally, we invoke the linearized kinematic conditions, $\partial_t \tilde{h}_\pm = \tilde{u}_z (k, z = \pm h_0/2,t)$, and obtain a set of coupled linear differential equations. The symmetric-antisymmetric decomposition, through $\tilde{h}_\textrm{sym} = \tilde{\delta h_+} - \tilde{\delta h_-}$ and $\tilde{h}_\textrm{anti} = \tilde{\delta h}_+ + \tilde{\delta h_-}$ [see Fig. \ref{fig:schem}(b)], appears as the natural modal decomposition for this system. These two modes relax independently to equilibrium, with distinct decay rates $\lambda_\textrm{sym}$ and $\lambda_\textrm{anti}$, since:
\begin{subequations}
\begin{equation}
\partial_t \tilde{h}_\textrm{sym} = -\frac{\gamma k }{\eta} \, \frac{\sinh^2(\frac{kh_0}{2})}{\sinh(kh_0) + kh_0} \tilde{h}_\textrm{sym} = -\lambda_\textrm{sym} \tilde{h}_\textrm{sym}\ ,
\label{eqn:eigen1}
\end{equation}
\begin{equation}
\partial_t \tilde{h}_\textrm{anti} = -\frac{\gamma k }{\eta} \, \frac{\cosh^2(\frac{kh_0}{2})}{\sinh(kh_0) - kh_0} \tilde{h}_\textrm{anti} = -\lambda_\textrm{anti} \tilde{h}_\textrm{anti}\ .
\label{eqn:eigen2}
\end{equation}
\end{subequations} 

\begin{figure}
\begin{center}
\includegraphics[width = 0.95 \columnwidth]{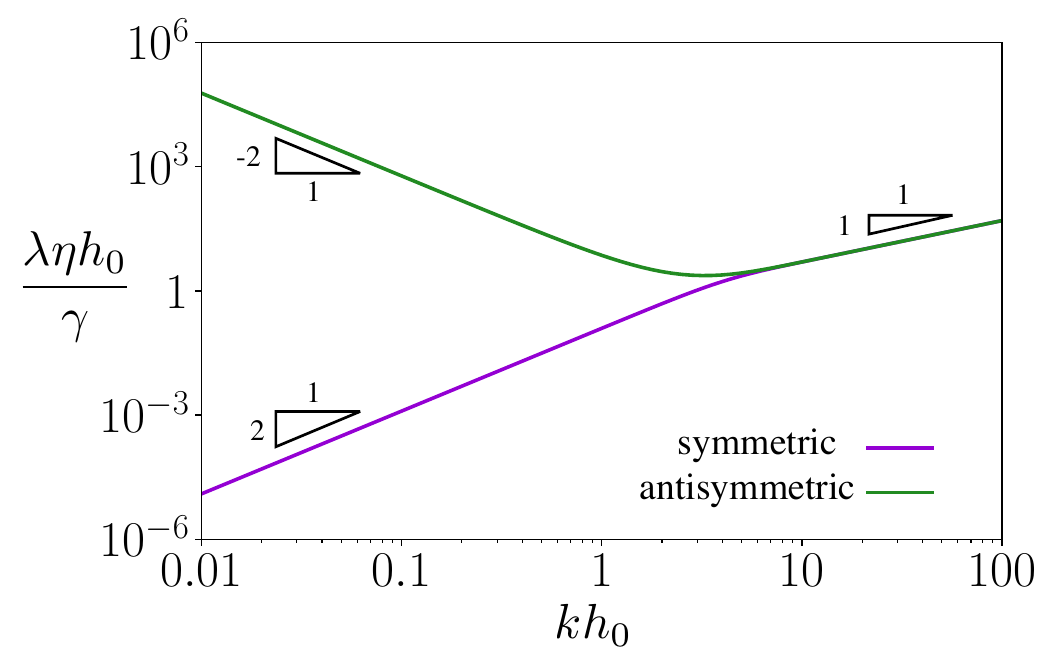}
\caption{Dimensionless decay rates of the symmetric and antisymmetric modes (Eqs.~(\ref{eqn:eigen1}) and~(\ref{eqn:eigen2})) as a function of the dimensionless wave number. The slope-triangles indicate power-law exponents.}
\label{DecayRate}
\end{center}
\end{figure}

The dimensionless decay rates are plotted in Fig.~\ref{DecayRate} as a function of the dimensionless wave number $kh_0$. For each rate, two asymptotic behaviors can be distinguished. At large $kh_0$, both rates exhibit the same limit: $\lim_{k \to \infty} \lambda(k) = \frac{\gamma k }{\eta}$. At small $kh_0$, the symmetric rate becomes identical to the one in the symmetric long-wave free-standing film model: $\lim_{k \to 0} \lambda_\textrm{sym} = \frac{\gamma h_0k^2}{8\eta}$ \citep{MI2016,erneux1993nonlinear}, and thus Eq.~\eqref{eqn:eigen1} reduces to a heat-like equation in Hankel space, with a diffusion coefficient $ \frac{\gamma h_0}{8\eta}$. In the same limit, the antisymmetric rate has a different scaling law: $\lim_{k \to 0} \lambda_\textrm{anti} = \frac{6 \gamma }{\eta h_0^3 k^2 }$. Therefore, long waves are quickly damped for the antisymmetric mode. We note that $\lambda_\textrm{anti}$ has a minimum at $k \simeq 3.28 /h_0$, corresponding to a slowest mode, which sets the relaxation dynamics.

The model relies on the assumption of a Newtonian fluid. As such, it must be compared to experimental profiles corresponding to annealing times longer than the polymeric relaxation time. Thus, we take the experimental profiles at $t_\textrm{exp} = 5$ min as the initial conditions for the model (see Fig. \ref{fig:profs_topbot}). Equations~\eqref{eqn:eigen1} and \eqref{eqn:eigen2} are then solved, yielding: 
\begin{equation}
\label{eq:profile_time}
\tilde{h}_\textrm{sym/anti} (k, t) = \tilde{h}_\textrm{sym/anti}(k,0) \, \exp\bigg[-\lambda_\textrm{sym/anti} (k) t\bigg]\ , 
\end{equation}
where $t = t_\textrm{exp} - 5$~min. 
\begin{figure}
    \includegraphics[width=1\columnwidth]{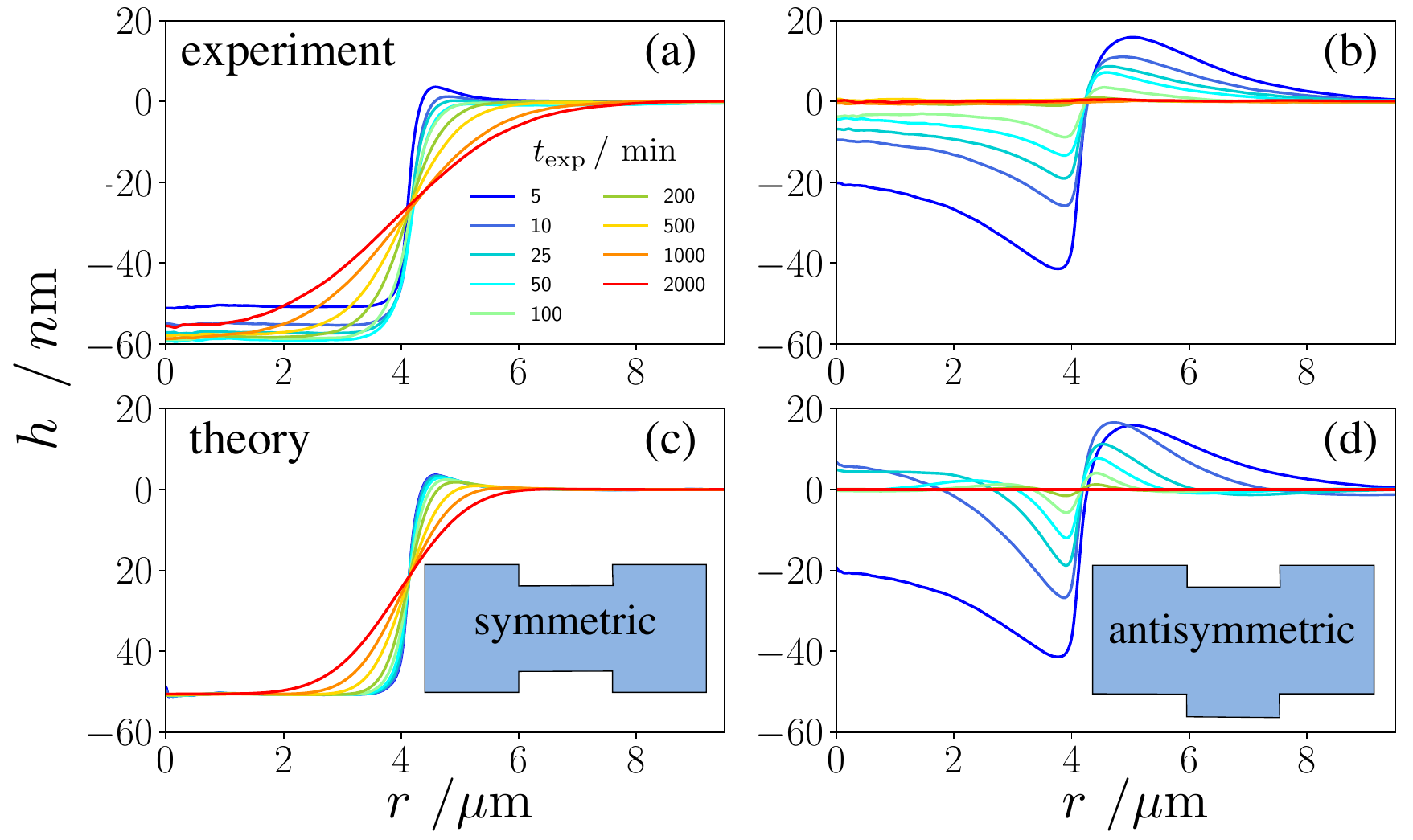}
      \caption{Symmetric (a) and antisymmetric (b) modes of the experimental (angular averaged) profiles for various times. The colors correspond to the same times as in Fig~\ref{fig:profs_topbot}. Symmetric (c) and antisymmetric (d) modes of the theoretical profiles, according to Eq.~\eqref{eq:profile_time}, for various times, and with the experimental profiles at $t_\textrm{exp} = 5$~min as the initial conditions ($t=0$).}
          \label{fig:sym}
 \end{figure} 
The symmetric and antisymmetric modes for the experimental and theoretical profiles are shown in Fig.~\ref{fig:sym}. There is a qualitative agreement between theory and experiments. Notably, the symmetric mode exhibits a self-similar behavior when plotted (not shown) as a function of the variable $(r - r_0)/t^{1/2}$. This result for free-standing films is to be compared to the capillary levelling of a cylindrical hole in a film supported on a substrate, that shows a self-similar behaviour in $(r - r_0)/t^{1/4}$~\cite{MB2014}. In contrast, the antisymmetric mode vanishes rapidly, on a time scale on the order of $\sim200$~min, meaning that the top and bottom interfacial profiles become perfectly mirror-symmetric, as observed in Fig. \ref{fig:profs_topbot}. The long waves appeared to be damped more quickly than the short ones, in agreement with the limiting scaling behaviors of $\lambda_\textrm{anti} (k)$ (see Fig.~\ref{DecayRate}).

\begin{figure}
\centering
\includegraphics[width=1\columnwidth]{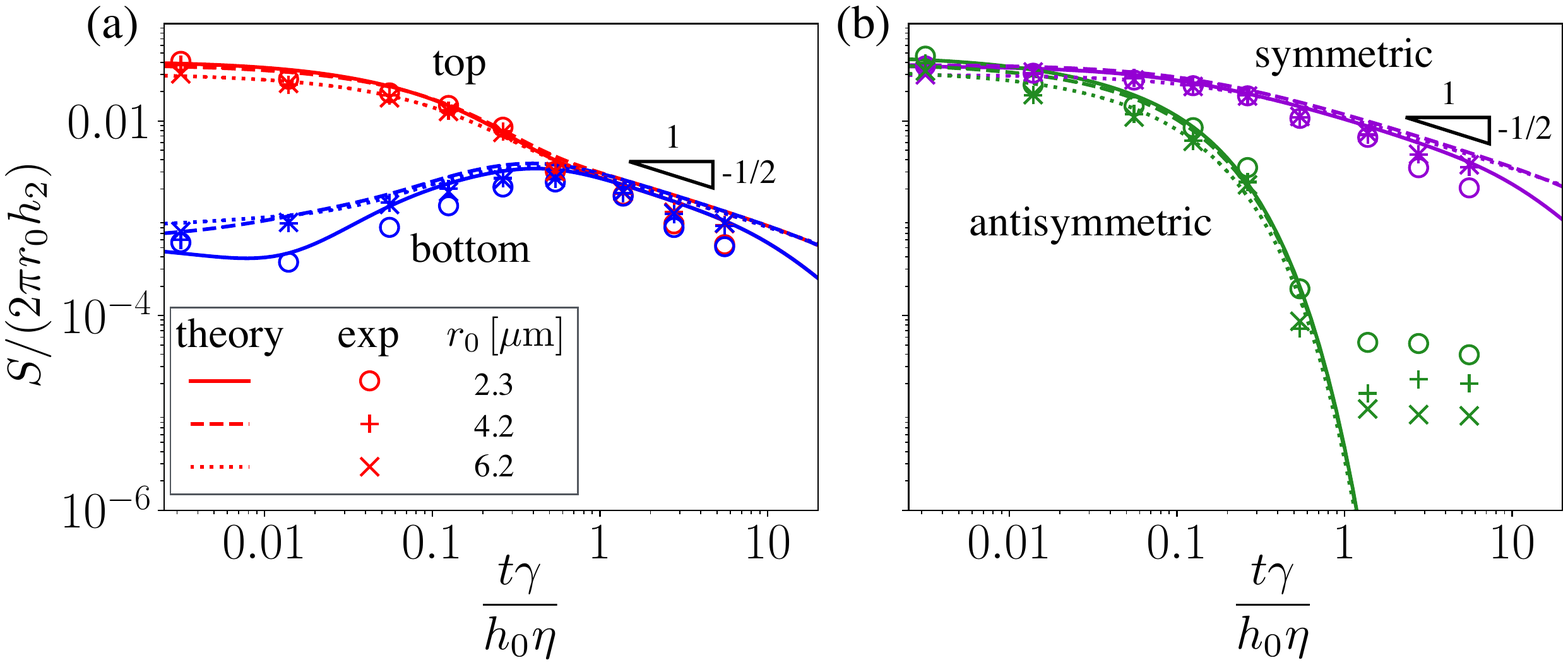}
\caption{Dimensionless excess surface area as a function of dimensionless time. The experimental data for three different holes are shown with different marker symbols, as indicated. The corresponding theoretical data are shown with different line styles, as indicated. (a) The top and bottom interfacial profiles.  (b) The symmetric and antisymmetric modes. The slope-triangles indicate power-law exponents.}
\label{fig:normarea}
\end{figure}
A measure of the distance to equilibrium lies in the excess capillary energy, which is proportional to the excess surface area with respect to a flat film, $S_i= 2\pi\int_0^{\infty} \textrm{d}r \,r (\sqrt{1 + (\partial_r h_i)^2} - 1) $, where the index $i$ can refer to $+$, $-$, $\textrm{sym}$, or $\textrm{anti}$, depending on the profile/mode in question. The excess surface area reduces to $S_i \simeq \pi\int_0^{\infty}\textrm{d}r \, r (\partial_r h_i)^2$ in the small-slope limit (valid at $t_\textrm{exp} > 5$ min). Figure~\ref{fig:normarea}(a) shows the excess surface areas of the top and bottom profiles, normalized by the initial excess surface area, as a function of dimensionless time, $\gamma t/(h_0\eta)$, for three holes of different initial radii, $r_0 = 2.3 \, \mu$m, $4.2 \, \mu$m, and $6.2 \, \mu$m on the same film. The trends are consistent with the intuitive expectations illustrated in Fig.~\ref{fig:schem}(c), and the theoretical curves are in excellent agreement with the experimental data, which validates the hydrodynamic model. We further see that the top interface, which has an initially high excess surface area, exchanges energy with the bottom one, causing the excess surface area of the latter to initially increase. This happens through viscous vertical flow, a process that continues until the formation of a mirror-symmetric interfacial profile on the bottom of the film, at $\gamma t/ (h_0\eta) \sim 0.5$, after which the excess surface areas of both interfaces are equal. At later times, the surface areas decrease following a power law $S \sim t^{-1/2}$ because of the self-similar properties of the heat-like equation that governs the symmetric mode.
 
With the modal decomposition above, one can also define and plot the symmetric and antisymmetric surface areas, $S_\textrm{sym}$ and $S_\textrm{anti}$ respectively, as functions of the dimensionless time (see Fig.~\ref{fig:normarea}(b)). The two modes relax with different dynamics. The symmetric mode exhibits a longterm $S_\textrm{sym} \sim t^{-1/2}$ scaling, as a result of lateral flow. In contrast, the vertical flow in the antisymmetric mode dissipates energy much more quickly, with a typical time scale $\sim \eta h_0 / \gamma$, that is identified as being the symmetrization time scale. 
The experiments reveal that this symmetrization time scale does not depend on the initial radius of the hole, and is set solely by the dynamics of the slowest relaxation mode, \textit{i.e.} the Fourier-Bessel mode $k$ at which $\lambda_{\textrm{anti}}(k)$ is minimal (see Fig.~\ref{DecayRate}).

It is interesting to note that in real space the governing equation of the antisymmetric mode is $\frac{1}{6}\eta h_0^3 \partial_t \nabla^2 h_\text{anti} = \gamma h_\text{anti}$, in the long-wave limit. Upon taking the Laplacian of this expression, we recover on the right-hand-side the Laplace pressure difference $\delta P = \gamma \nabla^2 h_\text{anti}$ across the film. Then, the mid-plane line $H = h_\textrm{anti}/2$ follows the equation $\frac{1}{3}\eta h_0^3 \, \nabla^4 \partial_t H = \delta P$. This equation corresponds to the torque balance in the liquid film~\cite{howell1996models, pfingstag2011linear, srinivasan2017wrinkling}, and is the viscous analogue of the F\"oppl-von K\'arm\'an equation for an incompressible elastic membrane in pure bending, where the bending modulus is replaced by $\eta h_0^3/3$, and the deflection field is replaced by the deflection rate $\partial_t H$. 

In conclusion, we have reported on the symmetrization dynamics of cylindrical holes in free-standing thin viscous polymer films. The topographies of both interfaces of the films were measured using AFM at various times, to track the evolution of the films while they were annealed above their glass-transition temperature. The films were found to undergo a rapid symmetrization process in order to equilibrate the Laplace pressures of the two liquid-air interfaces. This process transfers excess surface energy between the two interfaces, and eventually results in mirror-symmetric profiles on both sides of the film. A full-Stokes flow linear hydrodynamic model was developed to rationalize the observations. The model revealed the important roles of two modes, that differ by their symmetry with respect to the mid-plane of the film. The antisymmetric mode is associated with vertical flow, driven by the pressure gradient across the film, and exhibits faster dynamics than the symmetric mode, associated with lateral flow. The vertical symmetrization was found to occur on a universal time scale $\eta h_0 / \gamma$, while the symmetric mode dominates at later times. Surprisingly, the evolutions of the interfacial profiles, when decomposed into the symmetric and anti-symmetric components are found to obey power laws, with the decrease in surface area of the symmetric mode scaling as $t^{-1/2}$, analogous to the heat equation. 

We gratefully acknowledge financial support by the Natural Science and Engineering Research Council of Canada. We thank Hendrik Meyer for valuable discussions, and Paul Fowler for help with preliminary experiments.

\end{document}